\documentclass[useAMS]{mn2e}

\usepackage{graphicx}	
\usepackage{amsmath}	
\usepackage{amssymb}	
\usepackage{multicol}       
\usepackage{bm}		
\usepackage{pdflscape}	
\usepackage[utf8]{inputenc}
\usepackage[T1]{fontenc}
\usepackage{ae,aecompl}

\title[Confirmation of the wide binary anomaly.] {A recent confirmation of the wide binary gravitational anomaly.} 

\author[X. Hernandez, P. Kroupa] {X. Hernandez$^{1}$\thanks{xavier@astro.unam.mx},
  Pavel Kroupa$^{2,3}$\thanks{pkroupa@uni-bonn.de} \\ 
$^{1}$ Universidad Nacional Aut\'{o}noma de M\'{e}xico, Instituto de Astronom\'{\i}a, A. P. 70-264, 04510, CDMX, M\'{e}xico \\
$^{2}$ Helmholtz-Institut für Strahlen- und Kernphysik, Universität Bonn, Nussallee 14-16, 53115 Bonn, Germany. \\
$^{3}$ Astronomical Institute, Charles University, V Holesovickach 2, 18000 Praha, Czech Republic.\\ 
}

\date{Released 30/10/2024}

\begin{document}

\label{firstpage}

\maketitle

\begin{abstract}
  Concerning recent published studies exploring the presence or otherwise of a gravitational anomaly at low accelerations in wide binary stars
  as observed by the {\it Gaia} satellite, the paper published by Cookson on the subject last year presents an interesting case. In that study,
  RMS values of binned relative internal velocities in 1D for wide binaries are compared to Newtonian predictions for that quantity, with the
  author concluding that the data presented show no indication of any inconsistency with Newtonian expectations. However, the comparison presented
  is critically flawed, as the Newtonian predictions used refer to wide binaries with mean total masses of 2.0 $M_{\odot}$. This is larger than
  the 1.56 $M_{\odot}$ value which applies to the data used in said paper. In this short note we correct the error mentioned above and show that
  the data and error bars as given by Cookson are in fact inconsistent with Newtonian expectations. Contrary to the assertion in that study,
  the data presented there actually show a clear anomaly in the low acceleration gravitational regime, {  with the overall slope of the
  velocity-separation scaling being inconsistent with Newtonian expectations at a 3.3$\sigma$ level.} Based on the data presented in the paper
  by Cookson, wide binary systems show a clear Milgromian deviation from Newtonian dynamics.
\end{abstract}

\begin{keywords}
  gravitation --- stars: kinematics and dynamics --- binaries: general --- statistics
\end{keywords}

\section{Introduction}

Solar mass wide binary stars with separations above a few thousand au probe the low acceleration regime, $a \lesssim a_{\rm{0}}=1.2
\times 10^{-10} \rm{m s}^{-1}$, where gravitational anomalies at galactic and cosmological scales generally attributed to an as yet undetected
dark matter component occur. As first proposed by Hernandez et al. (2012), the distribution of relative internal velocities
between components of wide binaries will hence provide a critical test of gravity at the same acceleration regime where at much larger scales
dark matter is invoked to force an agreement between observations and general relativity (GR) and Newtonian gravity. If wide binary kinematics are
found to be in accordance with Newtonian predictions, then modified gravity proposals, which imply deviations from Newtonian dynamics in the low
acceleration regime, would be strongly disfavoured. On the other hand, if a gravitational anomaly is found in wide binaries, a low acceleration
validity limit for classical gravity will have been found, removing any astrophysical justification for the dark matter hypothesis, and evidencing
the need for the development of extensions to GR. Notice that independent constraints on the local density of dark matter imply a limit of
$\rho_{\rm{DM}}<0.01 \rm{M}_{\odot} \rm{pc}^{-3}$ (e.g. Read 2014), thus, any boost at even tenths of a percent level detected in the relative velocities
of wide binaries at separations of order 6 kau can not be explained through invoking dark matter as currently envisioned.

The release of the third {\it Gaia} catalogue has, for the first time, made available astrometric data of sufficient accuracy to undertake wide
binary gravity tests. As a consequence, over the last couple of years a series of published studies have appeared in this field. Given the
significant potential relevance of this test, all such works must be examined with care. Two independent groups using various sample selection
strategies and statistical tests have published numerous analysis showing consistent results for an accurate match to Newtonian expectations
in the high acceleration region for separations below 2000 au where Newtonian gravity, and the most well studied modified gravity theory of MOND
(Milgrom 1983), coincide. These studies include careful testing of the methods applied using synthetic samples and a diligent treatment of observational
errors, which determine the confidence intervals of the reported results. Interestingly, these studies also coincide in their results for the low
acceleration regime, but finding in this region a gravitational anomaly. This consists of relative internal velocities between the components of wide
binaries with projected separations above 3000 au showing a $22^{+10}_{-6} \%$ boost on Newtonian expectations, in consistency with long standing MOND
predictions. Examples of the above are Hernandez (2023), Chae (2023), Hernandez et al. (2024), Chae (2024a) and Chae (2024b).

A study concluding {\it Gaia} wide binaries show a 19$\sigma$ preference for the best-fit Newtonian model over the best-fit MOND
one also appeared, Banik et al. (2024). This last paper however, suffers from several major flaws, as recently described extensively in
Hernandez, Chae \& Aguayo-Ortiz (2024). Very briefly, in Banik et al. (2024) the sample begins at the regime transition found by the previous
two groups, 2000 au, and hence lacks a deep Newtonian region where both models being compared coincide, as a self-consistency check
on the entire procedure. Also, noise-free models are compared directly to noisy data taken as a unique observational template, in spite of the fact
that the observational errors present are comparable, and in some cases even larger, than the velocity binning used. Ignoring the full effects
of observational noise explains the exaggerated confidence intervals quoted by the authors, and also introduces a bias towards a Newtonian model.

The reason for this last point is that if one starts from a MOND reality and then adds noise, the low velocity edge of the velocity distribution will
shift towards lower values which are no longer accessible to noise-free MOND models. Newtonian noise-free models on the other hand, are inherently
shifted towards smaller velocity values, and hence can easily match the low velocity edge of a MOND reality which includes the presence of observational
noise. The opposite does not happen on the high velocity edge of the velocity distribution, because there the noise-free Newtonian models
of Banik et al. (2024) easily match the MOND plus noise extension through the inclusion of flybys and other kinematic contaminants which the
authors do not remove from the sample, but attempt to account for in the 7-dimensional fits performed, where 6 of the 7 parameters refer not to
gravity, but to the various kinematic contaminants their sample was not cleared of originally. 

More recently Cookson (2024) appeared, henceforth Coo24, building on the approach of Hernandez, Cookson \& Cortes (2022), and exploring the
robustness of the results to variations in one particular detail of the sample selection of Hernandez, Cookson \& Cortes (2022). After testing
several options, the author settles on what he considers an optimal data set, and presents it in comparison to Newtonian predictions, to conclude
no gravitational anomaly is present in the data. As it is easy to check, this conclusion is not only invalid, but indeed the opposite, as the data
in Coo24 were compared to a Newtonian prediction calibrated for binary stars having a mass of $2 \rm{M}_{\odot}$ and not the $1.56 \rm{M}_{\odot}$ which
corresponds to the average mass of the binaries in the Coo24 sample.

In this short paper we expose this obvious mistake, and show the same final
data from Coo24, but compared to the Newtonian prediction corresponding to the correct normalisation. This shifts the Newtonian model
downwards towards slightly smaller velocities, to reveal the same low acceleration gravitational anomaly reported by both the Hernandez and the
Chae groups. As a further consistency check, the high acceleration results of Coo24 appear consistent with Newtonian expectations once
the correct normalisation is included, while these binaries with separations below 2000 au appear below the Newtonian predictions in the original
Coo24 paper, a self-consistency check which {  was} missed by the author. Indeed, the Coo24 data show the
robustness of the results of the Hernandez group to the particular detail being explored, and hence become a valuable confirmation 
of the wide binary gravitational anomaly.

The following sections probe into the Coo24 study in detail, and make explicit the points mentioned above. Section (2) discusses
the variations in sample selection explored in that paper, and Section (3) looks into the details of the flawed comparison presented against
Newtonian predictions for binaries having different masses from the ones included in the study, amongst other design flaws
of that paper. Since Coo24 {  gives only a} qualitative comparison of the data presented to a Newtonian model, Section (4) presents
quantitative statistical comparisons of the data in Coo24 against Newtonian expectations of those data. Section (5) presents our conclusions.

\section{Cookson (2024) Sample Selection}

\begin{figure*}
    \includegraphics[height=7.0cm,width=8.5cm]{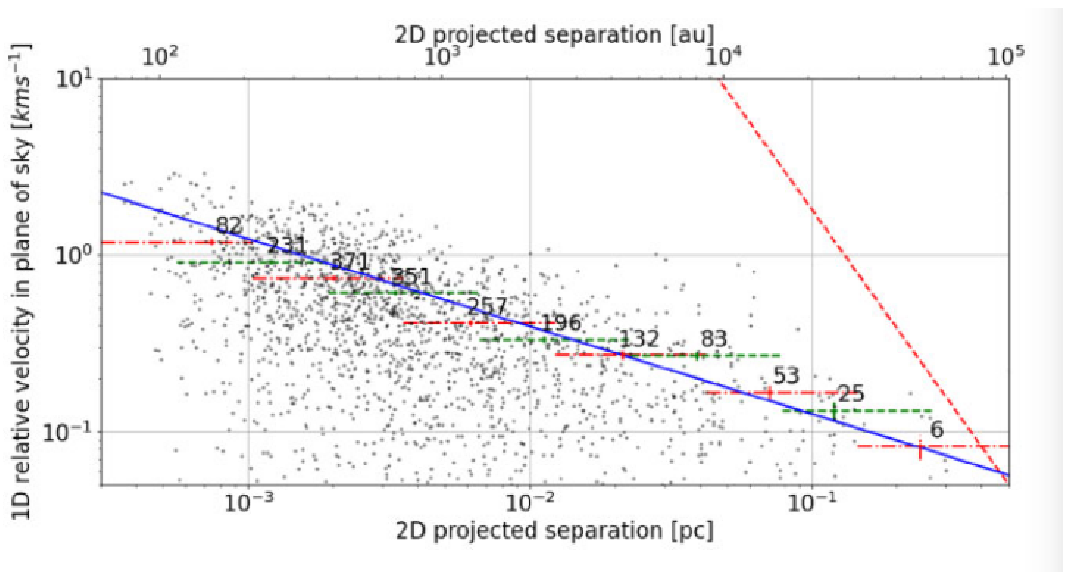}
    \hspace*{-6pt}
    \includegraphics[height=7.0cm,width=8.5cm]{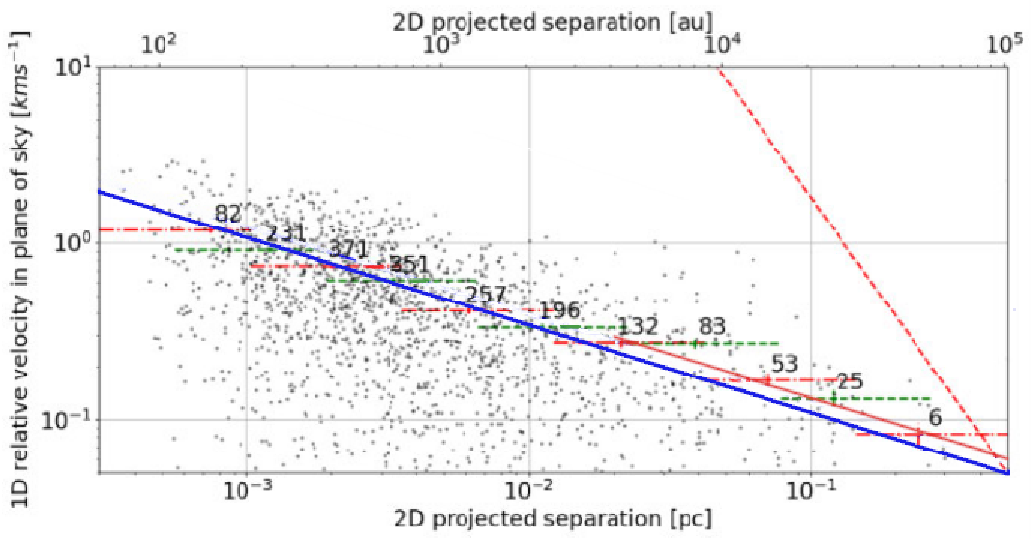}
    
    \caption{{\it Left}: The final figure in Cookson (2024), their fig.6. Each small black point represents a single wide binary, with the
      x-coordinate giving the  internal projected separation of the system, and the y-coordinate the 1D relative velocity between components
      of the binary. Each binary appears twice, for RA and Dec measurements. The red and green dashed lines give binned RMS values for velocities
      in the various sub-samples, for RA and Dec values, respectively, with the numbers above being the number of points per bin. The blue line
      gives the Newtonian prediction for the 1D RMS binned values of Jiang \& Tremaine (2010), calibrated for an assumed total binary mass of
      $2\rm{M}_{\odot}$. Aside from the clear anomalous velocity bin labelled '83', which {  Cookson (2024)} neglects to mention
      as such, the low acceleration region to the right of 3000au, 0.015 pc, shows binned {  values} consistent with the blue
      line, as they lie only $1\sigma$ above it, bins labelled '53' and '25'. The final point to the right is of little relevance, both on account
      of its poor statistical  value, and because it contains binaries in a regime where Galactic tidal effects and field stellar interactions
      become relevant, which all other groups working on the field for good reasons avoid, see text. What the author failed to notice was that in
      the high acceleration region to the left of 0.01 pc, which should serve as a consistency check of the overall procedure and a signal to any
      errors, the binned values actually lie below the blue line. The figure hence apparently shows a hitherto unknown high acceleration
      gravitational anomaly where observed binaries appear to rotate under a suppressed effective gravity. The reason for this lies in the fact that
      the average binary masses in the Cookson (2024) sample are actually not of $2\rm{M}_{\odot}$, but of $1.56\rm{M}_{\odot}$.
       {\it Right}: Same data as in the final figure in Cookson (2024), but showing this time the Newtonian prediction for binaries having the same
       mass as the average one present in the Cookson (2024) sample ($1.56\,\rm{M}_\odot$), blue line. The high acceleration region is now consistent
       with Newtonian
       predictions, and the low acceleration one shows again the clear close to 20\% velocity boost reported by the Hernandez and Chae groups, as evidenced
       by the solid red line of the MOND prediction. The slanted red dashed line in both panels is a hand fitted demarcation between bound binaries and
       flyby cases appearing in the original.
     }   
\end{figure*}

With the stated aim of exploring the influence of selection criteria on the conclusions drawn from wide binary gravity tests, Coo24
constructs various wide binary samples and compares their appearances on simple diagnostic internal separation vs. velocity plots.
The sample selection is based on what was used in Hernandez, Cookson \& Cortes (2022), allowing for modifications on one particular parameter.
The sample selection process begins with selecting all DR3 sources within a certain maximum
distance, in this case 143 pc, only marginally different from the value of 130 pc used in the publications of the Hernandez group. To this
first sample only very lax quality cuts are imposed, merely to have some confidence in that the sources selected are indeed stars. Then, a first
binary candidate list is produced by sampling 0.5 pc circles on the plane of the sky about each star in search of potential companions. In the
works by the Hernandez group, and following the ideas of El-Badry \& Rix (2018) and El-Badry et al. (2021), this step is optimised by requiring
that any candidate pair satisfies also that the distance between stars in the pair along the line of sight lies (to within their errors) within a
factor of order 2 of the separation on the plane of the sky of the pair. This last point was not included in the Coo24 study.

Next comes the step which is explored in the Coo24 paper, the removal of ambiguous candidate pairs. This, which has been termed degrouping,
consists of the removal of all candidate pairs which include any star which is also a member of any other candidate pair. At this point, an error
tolerance must be specified on the maximum parallax error below which a {\it Gaia} source will be considered as relevant for the degrouping process.
In Hernandez, Cookson \& Cortes, a value of 1\% was chosen at this point, given that the mean distance to the binaries studied was close to 100pc,
and maximum internal binary separations of 0.06 pc were considered, this implies that noisier excluded sources will have $1\sigma$ confidence intervals
for their distances to us along the line of sight of 1pc. Even if the reported distances of such sources coincide with those of target stars, this is already
at least 16.6 times larger than the largest binary separations considered, and hence refers to a source having a minimal probability of actually being a
perturber of any kind. Coo24 explores the effects of relaxing this $1\%$ maximum parallax tolerance for the degrouping phase of the
sample selection process.

Then, a series of quality cuts are applied to the data to remove flybys and cases where one or both stars in any binary might be binaries themselves,
through limits on the allowed internal {\it Gaia} single-star solution RUWE quality index, requiring a radial velocity measurement to be available for
each star and that the difference in this quantity between members of any binary be below 4 km s$^{-1}$ and a final filtering through the region of the
HR diagram identified by Belokurov et al. (2020) and Penoyre et al. (2020) to be least affected by unresolved binaries, which in this case would be
tertiaries. The details of the above quality cuts can be found in Hernandez, Cookson \& Cortes (2022) or Coo24.

Coo24 repeats the entire sample selection and astrometry processing procedures for various values of the maximum parallax tolerance error
to be considered in the degrouping phase, producing samples for $1, 1.4$ and $2.5\%$ tolerance errors. The results are presented in plots showing
the internal projected separation of each final binary considered against the relative velocity between components of each binary, both in RA and Dec.
Coo24 settles for the $2.5 \%$ parallax tolerance error during the degrouping phase as the optimal choice for this parameter. Results appear
in fig. 6 of that paper, which includes also a comparison against Newtonian predictions from the work of Jiang \& Tremaine (2010), which appear
as a blue solid line in this figure. This line is then compared against the RMS values of sub-samples binned into internal separation intervals. We
reproduce this figure in the left panel of Fig. 1.

Coo24 uses this figure to conclude that {\it Gaia} wide binaries show no indication of any low acceleration gravitational anomaly, as most of the
bins for separations larger than 0.01pc appear broadly consistent with the Newtonian line shown, with the exception of the one labelled '83', a {  fact}
Coo24 omits to mention. As it is crucial to understand exactly how the blue line in fig. 6 of Coo24 comes about, the following
section explores this point.

\section {Cookson (2024) Newtonian comparison}

In order to compare the points appearing in the diagnostic plots of figs. 4-6 in Coo24 against Newtonian expectations, the author presents binned RMS
values at a number of intervals. This is because the Newtonian expectation against which the data are compared are the results of the study by
Jiang \& Tremaine (2010). This last study simulates large populations of 50,000 wide binaries having a distribution of ellipticities and random
projection orientations, and evolved through numerical simulations over a 10 Gyr period, in the presence of the tidal field of the Milky Way at the
Solar Radius and encounters with molecular clouds. These binaries are then projected on the plane of the sky and the resulting distribution of points
is finally summarised in their fig. 7. This figure gives binned RMS values of the relative 1D velocity as a function of the internal projected separation
on the plane of the sky for the simulated binaries, precisely the quantities given in figs. 4-6 of Cookson (2024). However, fig. 7 in Jiang \& Tremaine
(2010) does not give separations or velocities in physical units, but in dimensionless quantities, so that the figure can be used to infer results for
binaries of any chosen total mass.

As it is trivial to check, one of us in Hernandez et al. (2012), assumed a 2$M_{\odot}$ total binary mass as a first order assumption to derive 
RMS internal 1D binary velocity values in km s$^{-1}$ vs. internal binary projected separations in pc from fig. 7 in Jiang \& Tremaine (2010). This
line in dimensional units first appeared in fig. 4 of Hernandez et al. (2012). In the interest of keeping a constant comparison line throughout the
various publications by the Hernandez group, the line was kept constant, assuming the same 2$M_{\odot}$ total binary mass. This was always made
explicit and borne in mind when comparing to data, e.g., the 12th paragraph in section (2) in Hernandez, Cookson \& Cortes (2022) makes it
explicit that a total binary mass of 2$\rm{M}_{\odot}$ was assumed towards arriving at the Newtonian comparison line for RMS velocities shown in the
figures where this appears in that paper. Also, the fifth paragraph in section (4) in Hernandez (2023) (which Coo24 cites) clearly mentions the value
of the total binary mass assumed in producing the RMS Newtonian expectation from the Jiang \& Tremaine (2010) results, with comments on the change
in normalisation between the data and the line being important included in the seventh paragraph of section (4) of that paper. That this line in
dimensional form has remained invariant from Hernandez et al. (2012) to Coo24 is trivial to check, as it is also to perform the scaling directly
from fig. 7 in Jiang \& Tremaine (2010) by assuming any desired total binary mass, in particular 2$\rm{M}_{\odot}$, to arrive at the blue line shown
in figs. 4-6 of Coo24.

The average binary mass in Hernandez (2023) was of 1.56$\rm{M}_{\odot}$, indeed, given the very similar sample selection used in Coo24, the average
binary mass in that work is also of 1.56$\rm{M}_{\odot}$ (Cookson, private communication). Thus, it is clear that the comparison presented in fig. 6
in Coo24 is critically flawed. Before correcting this obvious mistake, the original fig. 6 in Coo24 deserves a closer look, as shown in the left
panel of our Fig.1. The claim of no gravitational anomaly in the 2D projected separation $>$ 0.01 pc region, even when comparing against Newtonian
expectations for masses larger than the ones present, is somewhat suspect, not only because the '83' bin shows a clear velocity boost well above
the $1\sigma$ confidence interval given in the figure for that bin, but also because the following two bins appear above the Newtonian line,
although only at a $1\sigma$ level. Notice that the error bars are only the small vertical red and green lines, the horizontal red and green ones
simply denote the extent of the arbitrarily chosen separation bins. The final bin, containing only 6 binaries is irrelevant, not only on account of
the clear lack of statistical significance at such low occupancy numbers, but also because of the enormous projected separations covered of 0.5 pc.
Of the other groups having published on the topic, in both the Chae publications and in Banik et al. (2024), the samples are ended at 0.15pc,
Pittordis \& Sutherland (2023) stop at 0.1 pc, and, over an abundance of caution, in the Hernandez group publications the samples end at 0.06 pc.

It is clear that observed
projected separations are only lower limits on the present 3D separation between the components of observed binaries, which in turn, given the elliptical
orbits these systems present (Hwang et al. 2022), are only lower limits to the maximum orbital separations attained by a given binary. Given the
average interstellar separation of close to 1 pc for stars in the Solar Neighbourhood, all the groups listed above have chosen to stop the analysis
well before the maximum projected separation of 0.5pc considered in Coo24, to minimise the presence of nearby perturbers, present and past. {  A further
reason to stop a local wide binary gravity test at current projected separations below 0.5pc is that the Jacobi radius of a binary in the presence of
the tidal field of the Milky Way at the Solar Neighbourhood is 1.7 pc for a binary of a total mass of 2 solar masses, see for example
Jiang \& Tremaine (2010). For a total mass as that in the Cookson sample of 1.56 solar masses, the Jacobi radius will be 1.32 pc. }

A final important point to note regarding the original fig. 6 in Coo24, is the failure of the high acceleration consistency check. The inclusion of
a Newtonian region where there is no uncertainty as to what the result should be is intended as an internal check on the whole procedure. However,
as can be checked in the left panel in our Fig. 1, the binned RMS values of high acceleration binaries with separations below 0.01 pc actually
appear as sub-Newtonian. The original data presented in Coo24 thus display a clear gravitational anomaly, as high acceleration binaries show
smaller relative velocities than Newtonian expectations. {  This} should have alerted the author to the presence of an error in the comparison,
the mass normalisation mismatch noted above.

Finally, we reproduce in the right panel of Fig.1 the exact same data of the final fig. 6 in Coo24, but adjusting the Newtonian prediction to match
the average total binary mass of 1.56$\rm{M}_{\odot}$ present in the Coo24 sample. As can be seen, the high acceleration region for separations below
0.01 pc is now accurately consistent with the Newtonian prediction, with the exception of the tightest bin, where edge effects lead to a small offset.
The low acceleration region with separations above 0.01 pc on the other hand, now clearly shows a gravitational anomaly consistent with the
20\% velocity enhancement of MOND expectations, shown by the solid red line. At separations larger than 0.01 pc, the bins labelled '132', '53',
'25', and even the highly suspect final '6' bin, are all inconsistent with the blue line of the correct Newtonian prediction, and in excellent accordance
with the MOND prediction of the red line. The '83' bin is also inconsistent with Newtonian expectations, but remains anomalous when compared
to the MOND predictions, although by a smaller amount than what appears in the original fig.6 in Coo24. Going beyond this first qualitative
comparison would require a detailed statistical comparison against full velocity distribution predictions, of the type presented in e.g. Chae (2023)
or Hernandez et al. (2024).

In the conclusions of Coo24 both the results of the Hernandez and the Chae groups are dismissed on the basis of the mistaken diagnostic comparison
we reproduce in the left panel of Fig. 1, without including any detailed statistical test of the type appearing in e.g. Hernandez et al. (2024)
or in any of the publications of the Chae group, where careful comparisons of the full velocity distributions against simulated full velocity
distributions of Newtonian models are performed. Indeed, the sample selection parameter explored in Coo24 is not {  relevant to} the
publications of the Chae group, where the isolation of finally selected binaries is guaranteed through completely independent procedures. As is
evident from the right panel of our Fig. 1, what the results of Coo24 in fact show is exactly the same low acceleration gravitational anomaly which
has been reported in numerous independent analysis of the Hernandez and Chae groups, e.g. see fig. A1 in Hernandez (2023) or fig. 13 in Chae (2024a).

{ 
In the papers on this subject by the Hernandez group, prior to 2024, all wide binary kinematic comparisons
were performed against the 2 M$_{\odot}$ line in question, although the mean masses of binaries in the various samples treated
varied, typically at somewhat lower values. This was due to the fact that only
qualitative comparisons were attempted in those studies, mostly searching for any indication of a change
in regime, i.e. slope, in the data, aimed at identifying the region where any such change might be present.
Also, the intent was to leave the reference line fixed for comparison purposes between studies of that group,
the first of which was performed when mass estimates were still somewhat crude, for which reason the simple
2 M$_{\odot}$ reference value was originally adopted.

After Hernandez et al. (2023) clearly showed a change in the slope occurring at around
projected separations of 2000 au, the following paper, Hernandez et al. (2024), presented
a detailed statistical comparison against Newtonian predictions, not by comparing against the RMS line of
Jiang \& Tremaine (2010), but through careful comparisons of the full velocity distributions at both low
and high accelerations, against synthetic Newtonian velocity distributions using samples of upwards of
$10^{10}$ Newtonian simulated binaries including projection effects, ellipticity distributions and phase-occupancy
probabilities, through a formal Kolmogorov–Smirnov test. Through this procedure it was that the quantitative
estimate of the fractional gravitational anomaly quoted previously was calculated.
}

\section{Statistical comparisons of the Cookson (2024) data}
{ 
The assertion of no inconsistencies with Newtonian expectations in Coo24 where not only based on a flawed comparison of the data against
Newtonian predictions for masses larger than those present in the data analysed, but also based exclusively {  on a} qualitative
appraisal, as no statistical {  analysis} appears in Coo24.

In this section we perform simple linear regression fits to the reported
data points in Coo24, including their error bars. We use a model:

\begin{equation}
\rm{log}_{\rm{10}}(\Delta V_{\rm{RMS}}) = m \rm{log}_{\rm{10}}(s)+c,
\end{equation}  

\noindent where $\rm{log}_{\rm{10}}(s)$ and $\rm{log}_{\rm{10}}(\Delta V_{\rm{RMS}})$ are the x and y coordinates of the binned data points appearing
in fig.6 of Coo24, in order to obtain a qualitative comparison of these data to Newtonian expectations. For any range of projected separations
over which the mean mass of the binaries included remains close to constant, Newtonian expectations are of the $m=-0.5$ of Kepler's law. Indeed,
as checked in e.g. Hernandez et al. (2023), the mean mass across the range explored, for very similar binary selection criteria, does indeed vary
only marginally. The Newtonian prediction for the RMS relative sky projected velocity of the wide binaries included has to be computed numerically
by considering a large sampling including distributions of projection angles, phase occupancy and ellipticities, as done in Jiang \& Tremaine (2010),
from whose fig.7 one can derive $c_{\rm{2}}=-1.4$ for binary masses of $2 \rm{M}_{\odot}$.

We begin with the low acceleration $0.04< \rm{log}_{\rm{10}}(s/\rm{pc}) <0.5$ region, for
which one obtains a slope of $m=-0.550 \pm 0.0256$ and an intercept of $c=-1.404 \pm 0.0256$. These values compare satisfactorily to the
Newtonian expectation for the erroneous mass of $2 \rm{M}_{\odot}$ used in Coo24, of $m_{\rm{N}}=-0.5$ and $c_{\rm{2}}=-1.4$, leading to the claim in that
paper of no inconsistencies with Newtonian expectations in the results. Even before comparing to the correct mass of $1.56 \rm{M}_{\odot}$
relevant to the sample being studied, a consistency check is failed by the fully Newtonian interpretation, in that the high acceleration
$0.00055< \rm{log}_{\rm{10}}(s/\rm{pc})  <0.013$ region is clearly sub-Newtonian, with  $c=-1.476 \pm 0.020$. The inconsistency at this point is
at a level of 3.8 times the variance of the inferred value of the intercept in this region.

\begin{table*}
\begin{flushleft}
  \caption{Parameters for the linear regresion fts to fig. 6 of Cookson (2024).}
  \begin{tabular}{ | l | c c c c | }
  \hline
  \hline   
 Projected  &                                              &                     & $0.00055< \rm{log}_{\rm{10}}(s/\rm{pc}) <0.013$ \\
 separation &  $0.00055<\rm{log}_{\rm{10}}(s/\rm{pc})<0.013$ &  $0.04< \rm{log}_{\rm{10}}(s/\rm{pc}) <0.5$  &       and              \\
 range      &                                              &                     & $0.04< \rm{log}_{\rm{10}}(s/\rm{pc}) <0.5$      \\   
   \hline
 $m$        &  -0.501$\pm 0.003$      & -0.550$\pm 0.0256$  & -0.434$\pm 0.0004$     \\
 $c$        &  -1.476$\pm 0.020$      & -1.404$\pm 0.0256$  & -1.306$\pm 0.0022$     \\
 \hline 
 
\end{tabular} 

  Values for the parameters and their variances for fits to equation (1) of the data in three distinct projected separation, $s$, regions
  of fig. 6 of Cookson (2024). These can be compared to Newtonian expectations of $m_{\rm{N}}=-0.5$ for any relatively constant binary mass
  across a given interval, and $c_{\rm{2}}=-1.4$,  $c_{\rm{1.56}}=-1.454$ of Newtonian expectation for binaries of total masses of $2.0 \rm{M}_{\odot}$
  and  $1.56 \rm{M}_{\odot}$, respectively. While the low acceleration region is consistent with Newtonian expectations for $2.0 \rm{M}_{\odot}$, even
  comparing to this larger mass than the mean mass of the binaries in the sample ($1.56 \rm{M}_{\odot}$), the high acceleration region appears
  sub-Newtonian, while the overall fit has a slope which is inconsistent with Newtonian expectations. Comparing the high acceleration
  $0.00055< \rm{log}_{\rm{10}}(s/\rm{pc})  <0.013$ to the low acceleration  $0.04< \rm{log}_{\rm{10}}(s/\rm{pc}) <0.5$ regions, an $18\%$ velocity
  boost is apparent, in consistency with MONDian expectations.
\end{flushleft}
\end{table*}

\begin{figure}
    \includegraphics[height=7.0cm,width=8.5cm]{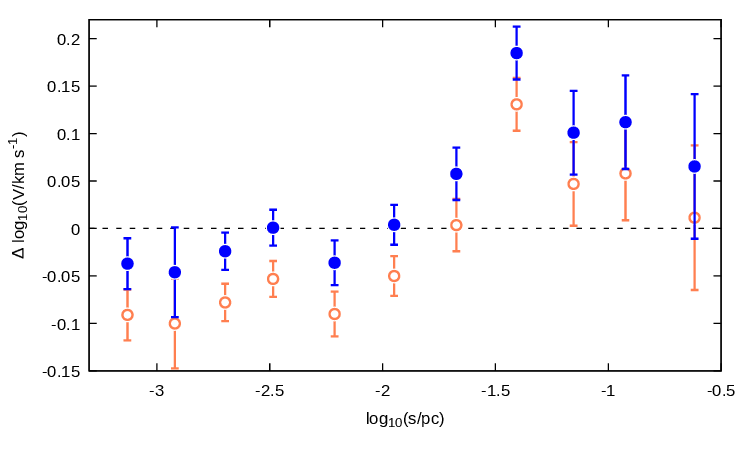}    
    
    \caption{The figure shows the distance between the data points and their $1\sigma$ confidence intervals from fig.6 in Coo24, and two Newtonian
      models: one corresponding to binary stars with a total mass of $2.0 \rm{M}_{\odot}$, and one corresponding to the same binary mass as the average mass
      of the binaries actually present in the Coo24 sample of $1.56 \rm{M}_{\odot}$, empty and solid symbols, respectively.
     }   
\end{figure}

Newtonian expectations for the correct total binary mass of $1.56 \rm{M}_{\odot}$ yield $c_{\rm{1.56}}=-1.454$, which now compares satisfactorily with the
results of Coo24 for the high acceleration region. However, the low acceleration region now clearly reveals a gravitational anomaly at a
level of two times the variance of the value of the intercept obtained in this region. If one fits the two above regions simultaneously,
one obtains a flatter than Newtonian slope at $m=-0.434\pm 0.0004$, inconsistent with Newtonian expectations at a level of over 100 times the
variance for this fitted quantity, a 3.3$\sigma$ inconsistency. The larger number of points and their small error bars, together with the flatter overall slope resulting
from fitting over two regions where the second one appears at a boosted amplitude, yield results completely inconsistent with Newtonian
expectations, regardless of the assumed binary mass. Thus, we see that in going from {  the} qualitative appreciation of a good Newtonian
fit in the low acceleration region presented in Coo24 to conclude a full consistency with Newtonian predictions, to a slightly more careful
statistical appraisal, shows the data presented in Co24 to be in fact very significantly inconsistent with Newtonian expectations. This would
have been obvious in Coo24 if an statistical measure of the (in)consistency of the high acceleration region to Newtonian expectations had
been included. In fact, even a superficial inspection of fig.6 in Coo24 shows the high acceleration region to be sub-Newtonian, revealing the
use of a larger than present mass in the Newtonian model used.

Once considering a Newtonian model of the correct mass, the data presented in fig. 6 of Coo24 show a very good fit to both Newtonian and MONDian
expectations in the high acceleration region (where both in fact coincide), and a good accordance with MONDian expectations in the low acceleration
region as well, where the expectations under that modified gravity theory are for a pseudo-Newtonian behaviour of  $m_{\rm{N}}=-0.5$, and a close to
$20 \%$ velocity boost. From the inferred intercepts in both regions we obtain a $18^{\rm{+12}}_{\rm{-5}} \%$ boost in going from the high to the low
acceleration regions, directly from the data in fig.6 of Coo24. Table 1 collects the results of the statistical fits presented above.

We have included only the clear Newtonian and clear pseudo-Newtonian $m\approx -0.5$ regions from fig.6 in Coo24 for two reasons: firstly,
the presence of the anomalous '83' bin in that figure. Notice that this bin is not evident in the previous figs. 4,5 in Coo24, and probably
reflects artefacts due to the excessive line-of-sight tolerance implied by the very large $2.5\%$ parallax error considered in the degrouping
process, perhaps in combination with the significant relaxation of colour signal-to-noise constraints present in Coo24 in comparison to those
of the Hernandez group, see section 2.2.2 in Coo24. Second, while the high acceleration limit of MOND is well defined (coinciding with
Newtonian predictions), and the low acceleration limit for local wide binaries is well defined as being pseudo-Newtonian in that the radial
dependence of the force is again $r^{-2}$, the precise normalisation in this low acceleration regime depends on
the details of the unknown transition function of MOND. More seriously, the details of the transition in terms of
wide binary relative velocities, are more dependent on the details of the problem, not only on the MOND
transition function, but also the full orbital dynamics of the binaries in question, which have only been
solved numerically under a series of approximations. There is hence no definitive MOND predictions for
exactly what to expect for the local wide binaries over the transition region, see for example, Chae \& Milgrom
(2022). For this reason, and not to mislead with any plotted curve over the highly uncertain
transition region, we prefer to show only the MOND expectations over the regions where the limit behaviour
is expected. For the above, and in the interest of keeping a fixed data set for the statistical comparisons presented,
we have excluded the transition region between the two ones where a clear $m\approx -0.5$ scaling is seen in fig.5 of Coo24.

We finish this section with Fig. 2, which complements the statistical fits previously mentioned. In it we plot the vertical distance between the points
reported in fig. 6 of Coo24 (which appear at the geometric mean of the x-coordinates of each bin, as given in Coo24) and two Newtonian models:
the one used in Coo24 corresponding to a mistaken mass of $2.0 \rm{M}_{\odot}$, empty dots, and the correct Newtonian model for those data assuming a
binary mass of $1.56 \rm{M}_{\odot}$, equal to the mean mass of the Coo24 sample, solid dots, together with the error bars on these data, taken also from
fig. 6 in Coo24.

When comparing to the erroneous mass model used in Coo24, the empty dots, we see than indeed, the low acceleration region reported as anomalous by the
Hernandez and Chae groups, the last three points to the right, appears consistent with the model; only two of these three points are marginally more
than $1\sigma$ away from the model. It is this which prompted the assertion in Coo24 of no evidence for any gravitational anomaly in the data presented.
However, an examination of the high acceleration region, the first 6 points to the left, reveals a very clear inconsistency, as all these points
appear as sub-Newtonian, in many cases beyond  $2\sigma$ and sometimes even $3\sigma$. The reason for this is of course, the use of an
erroneously large mass for the Newtonian model, as the mean mass of the binaries in the Coo24 sample is not $2.0 \rm{M}_{\odot}$ but $1.56 \rm{M}_{\odot}$.

Looking at the solid dots, which show the differences between the data in Coo24 and the predictions of the correct Newtonian model assuming a binary
mass of $1.56 \rm{M}_{\odot}$, we see that in the high acceleration region, the first 6 points in the plot, only 3 lie slightly more than $1\sigma$ away
from this model. This constitutes an internal consistency check of the overall procedure, which was {  missed} in Coo24. The low
acceleration region however, the last three points to the right, now shows a clear gravitational anomaly, the  $18^{\rm{+12}}_{\rm{-5}} \%$ velocity boost
described above. This will correspond to a $36 \%$ boost in the effective value of the gravitational constant, in complete consistency with
Milgromian gravitation.

}

\section{Conclusions}

We have shown that the study by Cookson (2024) suffers from a critical flaw which renders its conclusions void. Correcting the mistake in question,
and using the data presented in that paper, in fact reveals a clear gravitational anomaly at low accelerations, with relative velocities between
the components of observed binary stars being above Newtonian expectations.

{  Indeed, even when comparing against the mistaken $2\rm{M}_{\odot}$ Newtonian line used in Cookson (2024), a simple linear fit statistical analysis shows
  that the data presented in that paper are inconsistent with a Newtonian model, both in terms of the amplitude of the signal in the high
  acceleration regime and the overall slope of the full data range presented.
  }

Thus, the data presented in Cookson (2024) are in fact consistent with
recent studies showing a close to 20\% velocity boost for {\it Gaia} binaries with separations beyond 3000 au, e.g. Hernandez et al. (2023),
Chae (2023), Hernandez et al. (2024), Chae (2024a) and Chae (2024b), in accordance with generic MOND predictions of decades ago.

\section*{acknowledgements}

The authors thank an anonymous referee for a thorough report containing constructive criticism of our original manuscript, which led to
an enriched final version. Xavier Hernandez acknowledges financial assistance CONAHCYT and DGAPA grant IN-102624.

\section*{DATA AVAILABILITY}
All data used in this work will be shared on reasonable
request to the author.

\end{document}